\UseRawInputEncoding 
\documentclass[%
 reprint,
 amsmath,amssymb,
pra,
]{revtex4-1}

\usepackage{graphicx}
\usepackage{dcolumn}
\usepackage{bm}

\usepackage{xcolor}

\begin{document}

\preprint{APS/123-QED}

\title{Quantized Nonlinear Thouless Pumping}

\author{Marius J\"urgensen}
	\email{marius@psu.edu}
\author{Sebabrata Mukherjee}
\author{Mikael C. Rechtsman}
	\email{mcrworld@psu.edu}
 
\affiliation{%
 Department of Physics, The Pennsylvania State University, University Park, PA 16802, USA
}%



%

\date{\today}

\begin{abstract}

The sharply quantized transport observed in the integer quantum Hall effect can be explained via a simple one-dimensional model with a time-periodic, adiabatically varying potential in which electronic charge is pumped from one side of the system to the other. This so-called `Thouless pump' captures the topological physics of the quantum Hall effect using the notion of dimensional reduction: The time-varying potential mathematically maps onto a momentum coordinate in a conceptual second dimension. Importantly, this assumes an electronic system in equilibrium and in its ground state, that is, with uniformly filled bands below a Fermi energy. Here, we theoretically propose and experimentally demonstrate quantized nonlinear Thouless pumping of photons with a band that is decidedly not uniformly occupied. In our system, nonlinearity acts to quantize transport via soliton formation and spontaneous symmetry breaking bifurcations. Quantization follows from the fact that the instantaneous soliton solutions centered upon a given unit cell are identical after each pump cycle, up to translation invariance; this is an entirely different mechanism from traditional Thouless pumping of fermions in equilibrium. Our result shows that nonlinearity and interparticle interactions can induce quantized transport and topological behavior even where the linear limit does not.

\end{abstract}

\maketitle


\begin{figure*}
\includegraphics{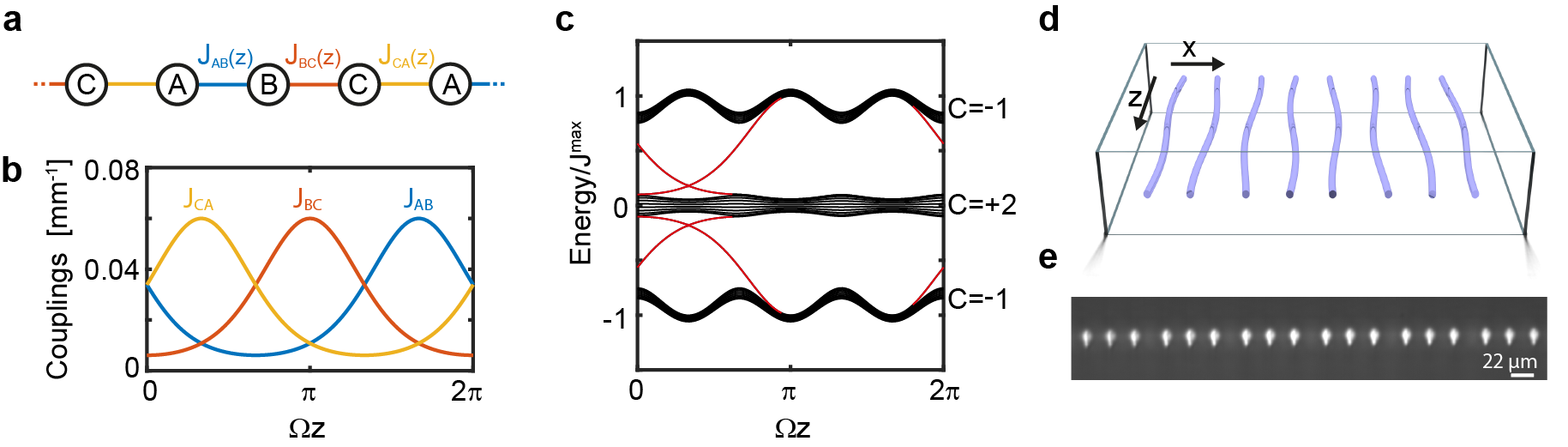}
\caption{\label{fig1} Photonic implementation of a topological Thouless pump. \textbf{a,} Schematic illustration of a Thouless pump model with three sites (A,B,C) per unit cell and $z$-dependent couplings between neighboring sites. \textbf{b,} Variation of the couplings during one driving period. \textbf{c,} Band structure calculated with open boundary conditions showing three bands with Chern numbers $C=\{-1,2,-1\}$. Red lines denote localized end states. \textbf{d,} Schematic illustration of the implementation of a pump model in a one-dimensional waveguide array where the waveguides are modulated in-plane. \textbf{e,} Micrograph of the output facet showing six out of ten unit cells for one lattice implementation.}
\end{figure*}

Soon after its discovery, the integer quantum Hall effect \cite{RN2444} was linked to a topological invariant, the Chern number \cite{RN2449}, which dictates the Hall conductance of a two-dimensional electron gas to a part in $10^{10}$ \cite{RN2496}. The Thouless pump came into being as a simplified model to understand this quantization of transport \cite{RN2448}: In a single cycle of the pump process, the number of electrons pumped from one side of the system to the other is precisely the Chern number of the occupied band - assuming a uniformly filled band. Beyond the condensed matter context of two-dimensional electron gases, topological behavior has been shown to be a general wave phenomenon, and has been observed in a variety of experimental platforms of fermionic and bosonic nature, including photonics \cite{RN2445,RN2446,RN42,RN31}, electronic circuits \cite{RN2355}, ultracold atoms \cite{RN2393,RN44,RN2461}, and exciton-polariton condensates \cite{RN2467,RN2468,RN2324}. Manifestations of Thouless pumping in particular have been observed in photonics \cite{RN2373,RN2483,RN2452,RN2470}, mechanical systems \cite{RN2471}, and ultracold atoms \cite{RN2469,RN2451}, including interacting pumps in the few particle limit \cite{RN2379,RN2495} or with half-filling \cite{RN2482}.
  
While there are similarities between the electronic and photonic cases, they have important differences: First, photonic topological systems are typically out of thermal equilibrium, and as such do not evenly populate any given bulk band; second, gain and loss are readily accessible in optical systems; third, interparticle interactions take on a fundamentally different form: At high optical intensities, photons can effectively interact with one another, mediated by an intensity dependent refractive index of the ambient medium. These differences have allowed for the observation of new physical effects not possible for electronic topological systems, including using non-Hermiticity to extract topological invariants \cite{RN2472}, topological band gap and edge solitons \cite{RN2453,RN2462}, nonlinearly-induced local topological transitions \cite{RN2473}, and other nonlinear/interacting effects \cite{RN2474}. Perhaps the most salient point of exploration in these systems is the interplay between topological physics and interparticle interactions and nonlinearity.

Here, we theoretically predict and experimentally observe quantized pumping of nonlinear eigenstates in interacting Thouless pumps based on optical waveguide arrays with Kerr nonlinearity. Importantly, the quantization is not rooted in a uniform occupation of a topologically nontrivial band as in traditional Thouless pumping \cite{RN2448}, but instead is of intrinsically bosonic nonlinear origin based on (1) the fact that the same nonlinear eigenstates are present at the beginning and end of each pump cycle (modulo translation invariance); and (2) the occurrence of spontaneous symmetry-breaking bifurcations. Hence, there exists no linear analogue of this effect in the domain of electronic Thouless pumps.


In the scalar-paraxial regime, the propagation of monochromatic light in a coupled waveguide array is governed by the discrete nonlinear Schr\"odinger equation:
\begin{equation}
\label{eq1}
	i \frac{\partial}{\partial z} \phi_n = \sum_{m}H_{n,m}^{\text{lin}}(z) \phi_m - g |\phi_n|^2 \phi_n.
\end{equation}
Here, $\phi_n(z)$ is the wavefunction (i.e. optical field) in waveguide $n$, $m$ and $n$ run over all waveguides, $H^{\text{lin}}$ is the linear $z$-dependent tight-binding Hamiltonian (e.g., describing a linear Thouless pump) and $z$ is the propagation distance, which for Thouless pumps plays the role of a synthetic wavevector dimension. The parameter $g$ describes the strength of the nonlinearity and is positive (negative) for a focusing (defocusing) Kerr nonlinearity. In the case of sufficiently low intensities, the equation reduces to the linear Schr\"odinger equation. From an experimental point of view, $g$ is dependent on the nonlinear refractive index coefficient of the underlying material, the effective area of the waveguide modes, and the wavelength. This nonlinear Schr\"odinger equation (with $g>0$) is equivalent to an attractive Gross-Pitaevskii equation describing bosonic interactions in a Bose-Einstein condensate in the mean-field limit. Indeed, the results we obtain below are generically applicable to a range of bosonic wave systems.

We use an off-diagonal implementation of the Aubry-Andr\'e-Harper (AAH) model \cite{RN2438,RN2486} with three sites per unit cell labeled as A, B and C (Fig. \ref{fig1}a). The off-diagonal AAH-model is described by a tight-binding Hamiltonian with equal on-site potential (which can be set to zero at all lattice sites) and real off-diagonal nearest neighbor couplings $J_n(z)$ that are periodic functions with a period of $\Omega$. The modulation of the couplings over one period is displayed in Fig. \ref{fig1}b, where the choice of colors corresponds to Fig. \ref{fig1}a. Replacing the general linear Hamiltonian in equation (\ref{eq1}) with the AAH-model results in:
\begin{equation}
\label{eq2}
	i \frac{\partial}{\partial z} \phi_n = -J_n(z) \phi_{n+1} - J_{n-1}(z) \phi_{n-1} - g |\phi_n|^2 \phi_n.
\end{equation}
This equation conserves the norm of the solution: $|\phi|^2 \equiv \sum_{n} |\phi_n|^2$. A pumping process (of the lowest band) can be understood intuitively using the following simplified description of the model: Suppose there is always only one coupling switched on, such that the intensity couples completely from one site to the next. Starting with an occupation of site A only, after switching on $J_{CA}$, the occupation shifts to site C. Subsequently, coupling $J_{BC}$ is turned on and finally $J_{AB}$ so that the wavefunction is pumped by one unit cell. This pumping behavior is topologically protected and stable against perturbations as long as the periodicity remains and the band gap is not closed. Therefore, when the model deviates from the described case of a perfect `hand-off' between one waveguide and the next, it still exhibits quantized pumping if a band is equally populated but shows a notable diffraction pattern.

To emphasize the relationship between a 1+1 dimensional pumping model (1 spatial and 1 propagation/temporal dimension) to a two-dimensional Chern insulator, we plot the band structure in Fig. \ref{fig1}c. At each point $z$ within a period, the energy eigenvalues of the instantaneous Hamiltonian (calculated for an array of 30 waveguides with open boundary conditions) are calculated and plotted. The band structure shows three bands with Chern numbers $C=\{-1,2,-1\}$ connected via topological end states (red). In contrast to two-dimensional Chern insulators, the topological end states appear and disappear during the evolution. A schematic illustration of a realization in a one-dimensional array of evanescently coupled waveguides is shown in Fig. \ref{fig1}d. The modulation of the coupling is achieved by periodically modulating the waveguides positions and therefore changing the spatial overlap of neighboring waveguide modes. The position of waveguide $n$ in the transverse direction is given by $x(z)=n \cdot d+\bar{d}\cos(2 \pi n/3 +\Omega z + \alpha_0)$ with $d$ being the average separation between two waveguides, $\bar{d}$ the spatial modulation strength and $\alpha_0$ an initial phase. The white light micrograph in Fig. \ref{fig1}e shows the output facet of a waveguide array with six out of ten unit cells.

\begin{figure*}
\includegraphics{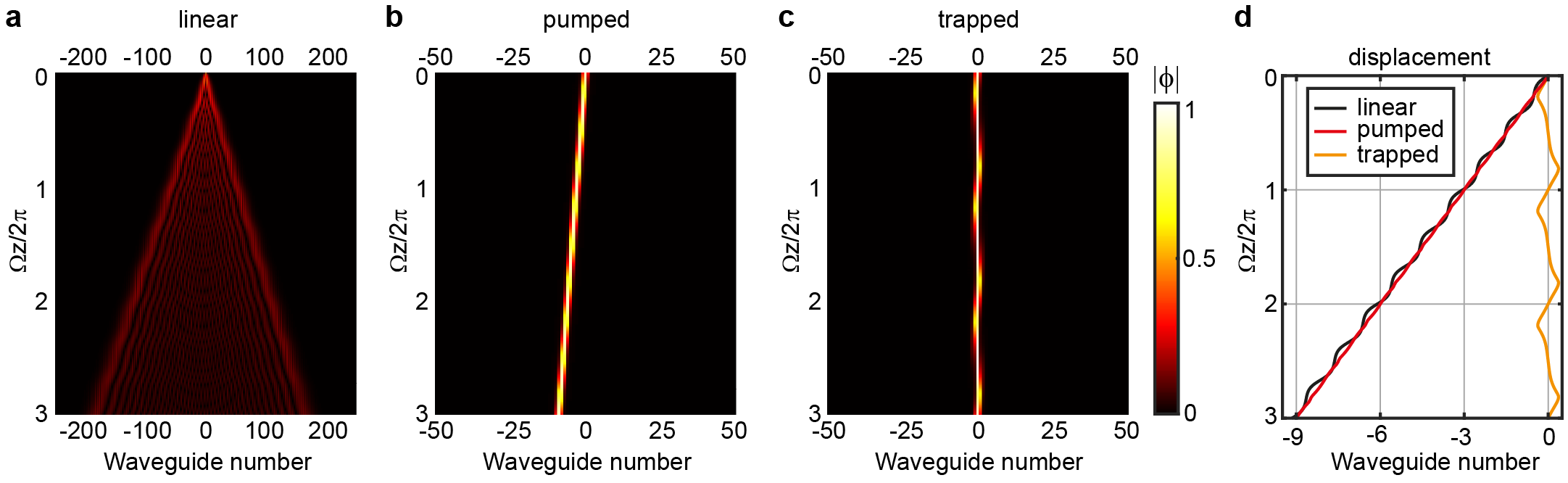}
\caption{\label{fig2} Linear and nonlinear propagation in topological Thouless pumps. \textbf{a,} Normalized amplitude of the wavefunction for a linear evolution over three periods for an input state with approximately uniform excitation of the lowest band. It develops a discrete diffraction pattern, while its center of mass is being pumped to the left by three unit cells. \textbf{b,} Nonlinear evolution for a pumped soliton with $g|\phi|^2/J^{\text{max}}=1.9$. The excitation is an instantaneous nonlinear eigenstate of the system. \textbf{c,} Same as \textbf{b} but with $g|\phi|^2/J^{\text{max}}=2.1$ and showing a trapped soliton. \textbf{d,} Displacement of the center of mass for the cases shown in \textbf{a-c}. The parameters for all figures are $d$=22 $\mu$m, $\bar{d}$=2 $\mu$m, $\alpha_0$=-2$\pi$/12 and $\Omega$=2$\pi$/$L$ with $L$=8550 mm. 
}
\end{figure*}

Using this Thouless pump, we demonstrate the differences between linear and nonlinear quantized pumping in three distinct regimes with different power: (1) A low power, linear regime in which the wavefunction evolves according to the linear Schr\"odinger equation; (2) An intermediate power regime in which we observe the formation of a soliton that is pumped by a fixed number of unit cells during a pump period; (3) A high power regime in which we observe a trapped soliton. We refer to the three regimes as the linear, pumped, and trapped regimes, respectively. Numerical propagation simulations of Eq. (\ref{eq2}) with a 4th order Runge-Kutta scheme are shown in Fig. \ref{fig2} for three periods. We use periodic boundary conditions; the number of waveguides exceeds the waveguides shown.

In the linear regime (Fig. \ref{fig2}a) the excitation is chosen such that the lowest band with Chern number $C=-1$ (see Fig. \ref{fig1}c) is approximately uniformly excited. This occupation is equivalent to a low temperature fermionic system with the Fermi level in the band gap, necessary for quantized pumping. Due to (spatial) diffraction, the wavefunction spreads during evolution, showing two dominant outer lobes similar to diffraction in a trivial array of straight waveguides. The bulk topological properties of the model are manifested in the transverse displacement of the center of mass by $C$ times the lattice constant (see also Fig. \ref{fig2}d).

Characteristic nonlinear behavior (at higher power) is displayed in Fig. \ref{fig2}b and c. In both cases, the excitation is a nonlinear eigenstate (i.e., a soliton \cite{RN2464}) of the instantaneous nonlinear Hamiltonian that bifurcates from the lowest band and is found using Newton's method (see Supplementary Information). The difference between the pumped regime (Fig. \ref{fig2}b) and the trapped regime (Fig. \ref{fig2}c), lies in the injected power in the system, which is $g|\phi|^2/J^{\text{max}}=1.9$ and $g|\phi|^2/J^{\text{max}}=2.1$, respectively. In both regimes, a suppression of spatial diffraction due to the focusing Kerr nonlinearity is observed. While the shape of the wavefunction changes during one period it remains strongly peaked and reproduces itself after each period, so that its shape resembles the shape of the input state. In the pumped regime (Fig. \ref{fig2}b), the wavefunction travels across the lattice with a displacement identical to the Chern number of the band from which it bifurcates ($C=-1$). After each period, the soliton is displaced by three lattice sites (one unit cell), which can be clearly seen in Fig. \ref{fig2}d. At high power (Fig. \ref{fig2}c), the wavefunction is trapped within the lattice: The soliton's center of mass oscillates around one lattice site (Fig. \ref{fig2}d), but ends up in the starting position after each cycle. The solutions shown in Fig. \ref{fig2}b and Fig. \ref{fig2}c are the pumped and trapped solitons, respectively.

The displacement of the wavefunction's center of mass in the linear case is well understood and arises from a uniform occupation of a topologically nontrivial band. The same principle is responsible for quantized charge transport in fermionic pumps even when disorder and interparticle interactions are present \cite{RN2437}. In stark contrast, here, the projection of the pumped soliton state onto the complete set of linear eigenstates does not show a uniform occupation of any band; indeed, the occupation is a strong function of power. Not surprisingly, propagation with the same input wavefunction but in the linear regime does not show quantized pumping. In the nonlinear regime we find numerically that the input soliton `tracks' the instantaneous localized stable soliton solutions at every $z$ during propagation. In other words, it behaves much as an eigenstate would in a {\it linear} time-varying, but adiabatic system, despite being a fundamentally {\it nonlinear} entity. Strictly speaking, these solitons are radiating due to the $z$-dependence of the Hamiltonian, but we find numerically that the radiated intensity is proportional to $\Omega^2$ (see Supplementary Information) and therefore becomes negligible for $\Omega \rightarrow 0$. This agrees with the results of Ref. \cite{RN2455}, proposing this as a nonlinear version of the adiabatic theorem. Thus, we can analyze the pumped and trapped solitons in terms of instantaneous nonlinear eigenstates.
 
\begin{figure*}
\includegraphics{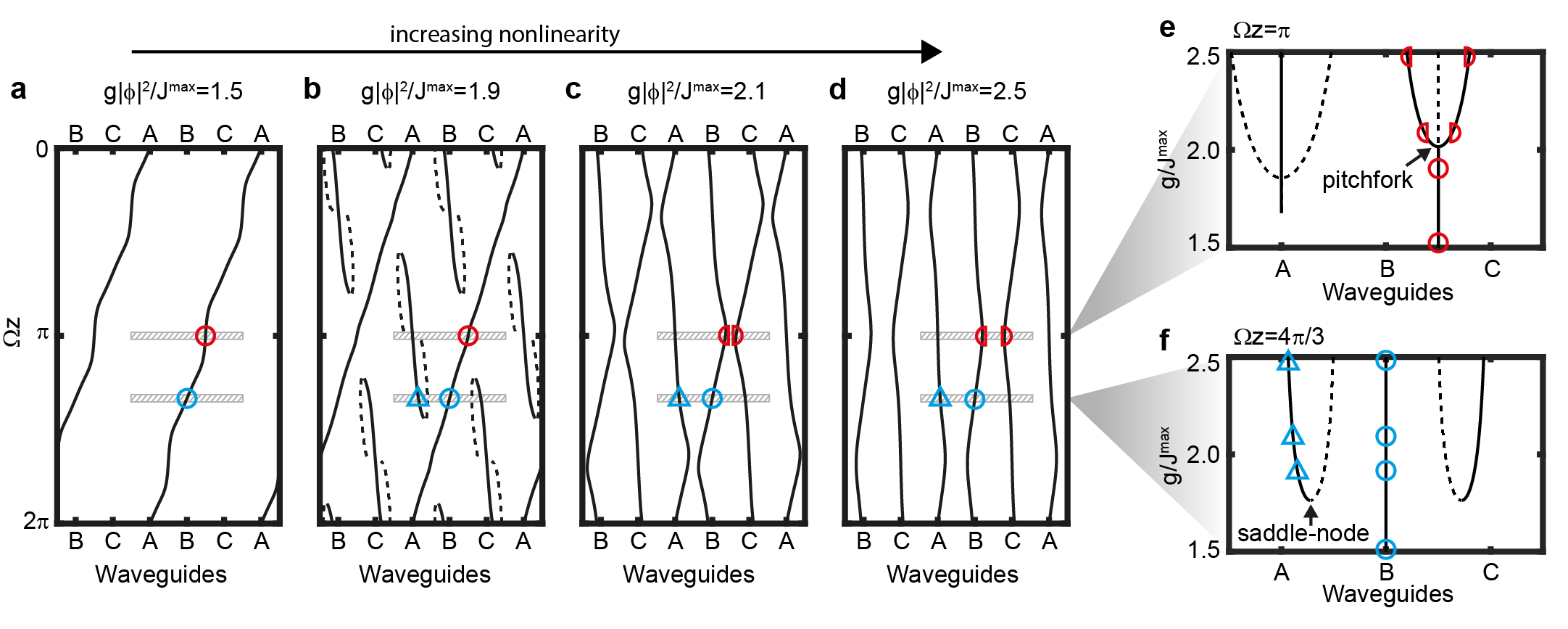}
\caption{\label{fig3} Mechanism of nonlinear pumping. \textbf{a-d,} Centers of mass of available soliton solutions at each value of $z$ in the pump cycle, showing contiguous paths. Black solid (dashed) lines indicate the position of the center of mass for stable (unstable) nonlinear eigenstates of the instantaneous Hamiltonian. \textbf{e-f,} Bifurcation diagrams for the nonlinear eigenstates at $\Omega z$=$\pi$ and $\Omega z$=4$\pi$/3, respectively, as a function of power. Blue and red symbols label specific soliton position of different branches which are shown in \textbf{a-d}. Each lattice consists of 30 sites and the parameters are the same as in Fig. \ref{fig2}.}
\end{figure*}

In order to explain the mechanism of pumping, we examine the position of the centers of mass of instantaneous solitons found at each time slice, $z$, as shown in Fig. 3a-d. We see that in Fig. 3a, in the pumped regime ($g|\phi|^2/J^{\text{max}}=1.5$), the soliton solutions follow a contiguous path through the lattice. For this power, we find one soliton solution (per unit cell), which is displaced along its path by one unit cell. For higher power ($g|\phi|^2/J^{\text{max}}=1.9$, Fig. 3b), new nonlinear eigenstates emerge (see for example also Ref. \cite{RN2488}) via saddle-node nonlinear bifurcations. (For a classification of bifurcations see for example Chapter 21 in Ref. \cite{RN2493}.) However, at this power, these bifurcations do not divert the original path of the soliton. In contrast, for still higher power ($g|\phi|^2/J^{\text{max}}=2.1, 2.5$, Fig. 3c,d respectively), a pitchfork bifurcation of nonlinear eigenstates, associated with a spontaneous symmetry breaking, gives rise to the splitting of the path of the soliton's center of mass, causing it to return to the site from which it started at the beginning of the cycle. A depiction of the centers of mass of the bifurcating solutions as a function of power are shown in Fig. 3e,f, and a clear animation of this process can be seen in the Supplementary Animation.

As we have seen, due to the periodic Hamiltonian, the nonlinear eigenstates at the beginning and end of each pump cycle are identical, and, due to translation invariance, exist for each unit cell. During adiabatic evolution, the soliton then tracks these eigenstates and when coming back to the beginning of a pump cycle the soliton is forced to occupy the initial state either in the same unit cell or displaced by an integer number of unit cells. In our model the Hamiltonian takes on the same form also after 1/3 (2/3) of one full period including a translation by one (two) sites. This allows us to observe quantized pumping also of single sites in addition to integer unit cells. While we focus here on the soliton that bifurcates from the lowest energy band that has a Chern number of -1, we show in the Supplementary Information that solitons that bifurcate from the two other bands are also pumped by the Chern number of the band from which they bifurcate. Indeed, in a range of models tested thus far, for which a contiguous path exists, the number of unit cells pumped at a suitable degree of nonlinearity (pumped regime) is the Chern number of the band from which the solitons bifurcate. We note further that the quantization of the soliton motion persists even at lower powers than those shown in Fig. 1 and 2, where the soliton is supported on many sites (see Supplementary Information).

Although no topological invariants are known for nonlinear systems, we can a-posteriori define a topological invariant for nonlinear pumping. To that end, we define an extended unit cell such that the exponential tails of the localized soliton become negligible within it. We then solve for periodic evolution of the soliton therein and include the potential induced by the nonlinearity as a linear potential in the Hamiltonian. With this, we are able to calculate the Chern number for the band describing the soliton evolution, which is $C=-1$ for the case shown in Fig. \ref{fig3}a (for details see Supplementary Information).

\begin{figure*}
\includegraphics{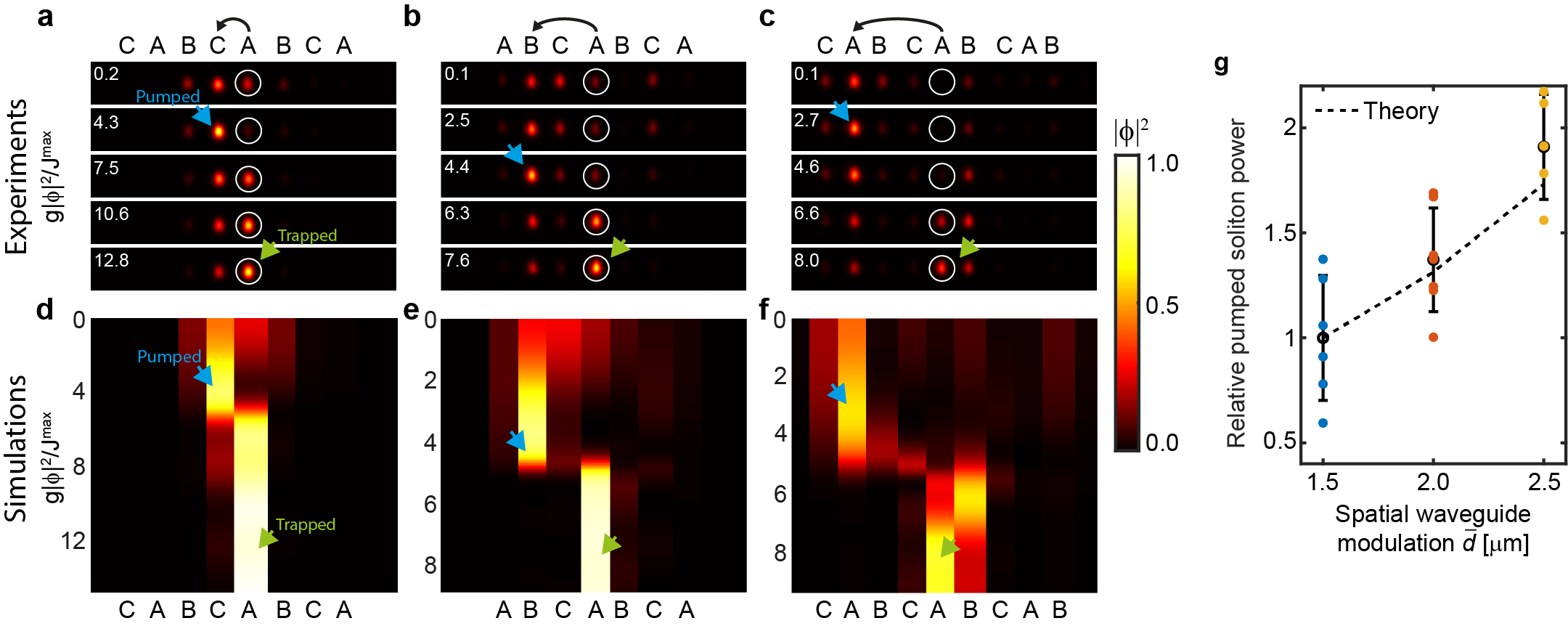}
\caption{\label{fig4} Experimental observation of quantized nonlinear topological pumping. \textbf{a,} Experimentally-observed normalized intensity pattern at the output facet of a 76\,mm long waveguide array ($d$=24 $\mu$m, $\bar{d}$=2 $\mu$m) after $1/3$ of a full period. White circles denote the excited waveguide; input power increases from top to bottom. Blue arrow denotes the observation of the pumped soliton and green arrow shows the trapped soliton. \textbf{b,} Same as \textbf{a} except for an array covering $2/3$ of a full pumping period ($d$=23 $\mu$m, $\bar{d}$=2 $\mu$m). \textbf{c,} Same as \textbf{a} except for an array covering one full pumping period ($d$=21 $\mu$m, $\bar{d}$=1 $\mu$m). \textbf{d-f,} Corresponding nonlinear tight-binding simulations for \textbf{a-c}, respectively. \textbf{g,} Input power (relative to the mean input power at modulation of $\bar{d}$=1.5 $\mu$m) required for maximum relative intensity in the pumped soliton as a function of spatial modulation strength $\bar{d}$ of the waveguides. Dots in color are measurements, black lines are the respective mean values with one standard deviation. The black dotted line shows the relative power value for the pitchfork bifurcation point.}
\end{figure*}


We experimentally probe the three regimes (linear, pumped, trapped) in a one-dimensional laser-written \cite{RN2463,RN2397} array of evanescently coupled waveguides with a focusing Kerr nonlinearity, which is present due to the ambient borosilicate glass. To reach the necessary degree of nonlinearity, we launch intense laser pulses into single waveguides, largely exciting the solitons. To reduce additional, unwanted nonlinear effects, especially the generation of new wavelengths via self-phase modulation, the pulses are temporally stretched to 2\,ps and down-chirped, accordingly. As a result, 76\% of the pulse intensity (equivalent to FWHM of a Gaussian) is found in a range of 20\,nm for input powers up to 6\,mW (see also \cite{RN2453,RN2462}). The wavelength-dependent change in the couplings is therefore similar to the intrinsic coupling strength uncertainty $\Delta J/J \approx \pm 10\%$ (see Supplementary Information). For the maximum propagation length of the waveguide arrays used in our experiments, chromatic dispersion effects can be neglected. The losses in the waveguides are $\approx$0.7 dB/cm and independent of power. Under these conditions, we can approximate the dynamics in our system with the nonlinear Schr\"odinger equation (see Eq. (\ref{eq2})).

In separate experiments, we observe nonlinear soliton pumping by one, two and three sites. Fig. \ref{fig4}a-c show the observed waveguide occupancies at the output facet. We use optical input powers of $\langle P \rangle_t=\{$0.1,2.0,3.5,5.0,6.0$\}$ mW, which we convert into $g|\phi|^2/J^{\text{max}}$ for each individual experiment with $g$=(0.07 $\pm$ 0.01)/mm per mW (time-averaged) input power (see Supplementary Information).
For low power ($g|\phi|^2/J^{\text{max}}$= 0.1 and 0.2), linear diffraction is observed and the intensity spreads over several sites. The displacement of the center of mass is not quantized, as the single-site excitation does not uniformly populate a band. For increasing input power, we observe strong localization of the wavepacket due to soliton formation (blue arrow): This is the pumped regime. A further increase in the input power causes light to strongly localize in the waveguide in which it was injected: This is the trapped regime. This transition from linear diffraction to a pumped soliton and finally to a trapped soliton is the experimental signature of quantized nonlinear pumping. For the corresponding simulations in Fig. \ref{fig4}d-f, we scaled the coupling function to the linear propagation in the individual array and the simulations clearly agree with the measured behavior. The lower contrast in the experiment arises from the fact that the tails of the pulse (in time) propagate linearly and diffract while only the region of high power in the temporal center of the pulse is affected by strong nonlinearity \cite{RN2453}. Nonetheless, the pumped soliton and a transition to a trapped soliton is clearly observed in simulation and experiment.

Finally, we analyze the prediction that the trapping is due to a pitchfork bifurcation (Fig. \ref{fig3}e). In Fig. \ref{fig4}g, we plot the experimentally-obtained optical input power required to observe the maximum relative intensity in the pumped soliton, which indicates the onset of the trapped regime, as a function of the degree of waveguide modulation, $\bar{d}$. We compare this to the numerically-obtained power for which the pitchfork bifurcation occurs. To directly compare the two, we normalize the power to unity at $\bar{d}=1.5\,\mu m$; on this basis, clear agreement between theory and simulation is observed. 

In conclusion, we have theoretically predicted and experimentally observed quantized nonlinear Thouless pumping followed by a transition to a trapped soliton at higher power. For a range of parameters and pump models, we numerically found soliton solutions which bifurcate from well-separated energy bands with Chern number $C$ and are pumped by precisely $C$ unit cells. Crucially, quantization is not based on uniform occupation of any topological band. Instead, exact quantization relies on the fact that the soliton solutions are identical at the beginning and end of each pump cycle up to a translation by a unit cell. We found that nonlinear bifurcations, splitting the propagation path, were responsible for the transition from pumped to trapped solitons. Our results extend the theoretical and experimental understanding of the interplay between nonlinearity and topologically nontrivial bands. Future directions of interest will be quantized nonlinear pumping in higher-order topological systems, as well as a complete understanding of the dynamics of Thouless pumping in systems with interactions on the few-particle level.


\section*{Acknowledgements}
We acknowledge fruitful discussions with Alexander Cerjan, Sarang Gopalakrishnan, Daniel Leykam, Oded Zilberberg and Panayotis Kevrekidis. We further acknowledge the support of the ONR YIP program under award number N00014-18-1-2595, ONR-MURI program N00014-20-1-2325, as well as by the Packard Foundation fellowship, number 2017-66821. M.J. acknowledges the support of the Verne M. Willaman Distinguished Graduate Fellowship at the Pennsylvania State University. Some numerical calculations were performed on the Pennsylvania State University's Institute for Computational and Data Sciences' Roar supercomputer.


\bibliography{bibliography}

\end{document}


\title{Supplementary Information\\for `Quantized Nonlinear Thouless Pumping'}

\maketitle

\section{Linear waveguide characterization}
\begin{figure}[h!]
\includegraphics{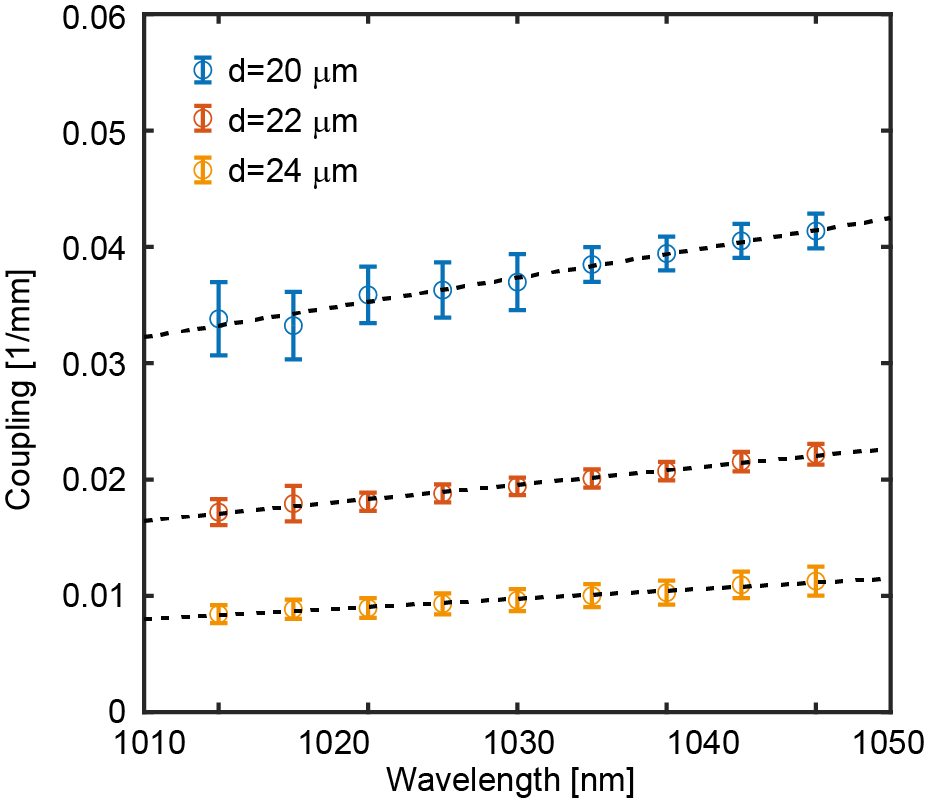}
\caption{\label{Supp:fig1} Characterization of the coupling strength. Coupling constants $J(d,\lambda)$ for wavelength $\lambda$, extracted from waveguide lattices with straight waveguides and equal separation,  $d$, between the waveguides. The errorbars show one standard deviation.}
\end{figure}

We measure the coupling strength $J(d,\lambda)$ as a function of separation, $d$, between waveguides and wavelength, $\lambda$. The waveguide arrays are characterized using a commercial supercontinuum source (NKT SuperK COMPACT) with a subsequent filter (NKT SuperK SELECT) to select the desired excitation wavelength. The light is focused into single waveguides in three separate arrays of straight waveguides ($d$=20 $\mu$m, 22 $\mu$m, 24 $\mu$m) with 30 waveguides each. To avoid edge effects, we only focus into the central 22 waveguides for $d$=22 $\mu$m, 24 $\mu$m and the central 12 waveguides for $d$=20 $\mu$m. The experimentally obtained output intensity pattern for each excitation is fitted with the theoretical output pattern in waveguide $n$, given by $\phi_n(L)=(i)^n \phi_0(z=0) B_n(2JL)$ via least-square fitting. Here, $L$ is the sample length of 76 mm, $B_n$ is the Bessel function of first kind of order $n$, $\phi_n$ is the amplitude in waveguide $n$ and $\phi_0(z=0)$ is the amplitude of the excited waveguide ($n$=0) at $z=0$. Fig. \ref{Supp:fig1} shows the mean coupling constants with one standard deviation. With increasing wavelength the optical waveguide mode width increases, leading to a larger overlap between neighboring modes and therefore higher coupling constants. The linear fits show that within a wavelength range of $\pm$10 nm around a central wavelength of $\lambda_0$=1030 nm the relative change $\Delta J$ in the coupling constant is $\Delta J= \pm 5 \%, \pm 6\%, \pm7 \%$ for $d$=20 $\mu$m, 22 $\mu$m, 24 $\mu$m, respectively. This change is comparable to the relative coupling uncertainty of about 7\%, 4\% and 10\% at the central wavelength of $\lambda_0$=1030 nm. At 1030 nm, fitting an exponential function, we obtain $J(d,\lambda_0) = 27.21 \exp(-330 s) /\text{mm}$, with $d$ in mm. In the simulations in the main text (Fig. 4d-f), we account for a remaining uncertainty in the coupling by using an additional multiplicative fitting factor for the coupling function, which is evaluated by fitting the linear output intensity pattern. The factors are 1.06, 1.28, and 1.12, for Fig. 4d-f respectively.

\section{Nonlinear waveguide characterization}
\begin{figure*}[h!]
\includegraphics{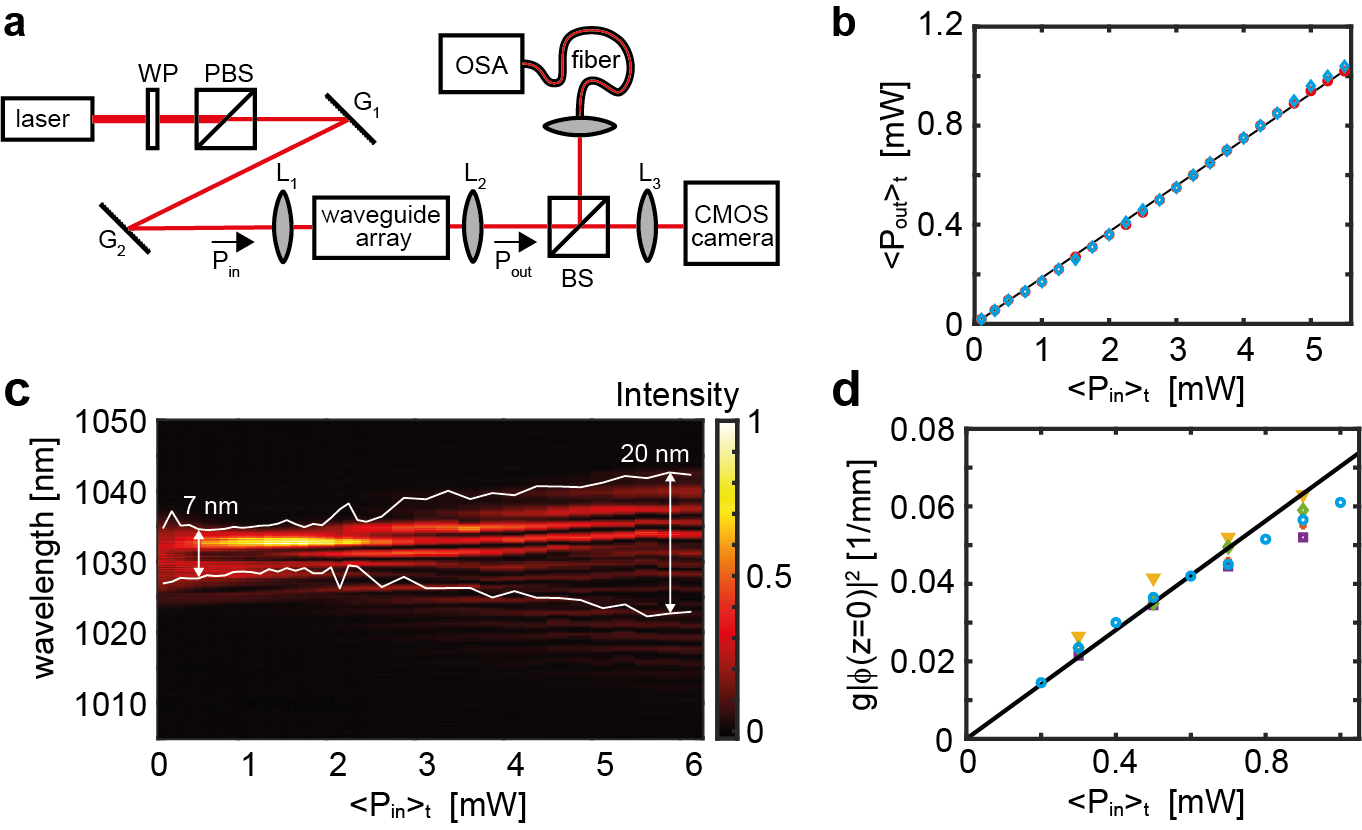}
\caption{\label{Supp:fig2} Nonlinear waveguide characterization. \textbf{a,} Simplified, schematic illustration of the experimental setup, including a half-waveplate (WP) together with a polarizing beam splitter (PBS) to adjust the power of the emitted laser pulses. Two gratings (G$_1$, G$_2$) temporally stretch the pulse to 2 ps. Lense L1 focuses the pulses into single waveguides within the waveguide array. Lenses L2 together with L3 image the output facet onto a CMOS camera. Simultaneously, using a further beamsplitter (BS), the light is additionally coupled into a fiber and its spectrum analyzed with an optical spectrum analyzer (OSA). \textbf{b,} Measured input to output power dependence for two data sets showing no nonlinear losses due to multi-photon absorption. The black line indicates a linear fit. \textbf{c,} Spectral distribution of the pulse after propagation of 76 mm in a lattice of straight waveguides with equal separation of 24 $\mu$m. The white lines denote the spectral range in which 76\% of the intensity (equivalent to FWHM of a Gaussian) are found. \textbf{d,} Theoretical nonlinear parameter $g|\phi(z=0)|^2$ versus the experimental time averaged input power. Depicted are values from five different waveguides in a lattice of straight waveguides with a separation of 24 $\mu$m. The black line indicates the value of $g$=(0.07 $\pm$ 0.01)/mm per mW (time-averaged) input power as the mean value with one standard deviation.}
\end{figure*}

The simplified schematic of the experimental setup used to carry out the nonlinear experiments is depicted in Fig. \ref{Supp:fig2}a. A commercial Yb-doped fiber laser (Menlo BlueCut) emits pulse trains at 5 kHz repetition rate (tunable). The pulses have a temporal width of approximately 260 fs and are spectrally centered at 1030 nm. We adjust the power using a half-wave plate (WP) together with a polarizing beam splitter (PBS). The parallel gratings G$_1$ and G$_2$ (Thorlabs GR25-0610) down-chirp and temporally stretch the pulses to 2 ps, see also Ref. \cite{RN2453}. The 2 ps pulses are focused into the single waveguides using lense L$_1$ (Thorlabs C280 TMD-B). Lenses L$_2$ (Thorlabs AC064-015-B-ML) and L$_3$ (Thorlabs LB1811-B-ML) are used to image the output facet onto a CMOS camera (ThorLabs DCC1545M), while the light is simultaneously injected into a fiber-coupled optical spectrum analyzer (OSA; Anritsu MS9740 A). The simplified schematic neglects additional mirrors and neutral density filters in the setup. We use a photodiode power sensor (Thorlabs S120C) to measure the time-averaged input and output power before L$_1$ and after L$_2$, respectively. Fig. \ref{Supp:fig2}b shows the observed dependence between input and output power for two single, separated waveguides. The linear relationship, visible via the linear fit, suggests that no nonlinear losses, e.g., via multi-photon absorption, occur, and therefore such losses can be neglected in our experiments. 

The pulses are temporally stretched in order to avoid a significant generation of new wavelengths via self-phase-modulation (SPM) in the sample. This is essential; if the spectrum is too broad, then the coupling constants between waveguides are not well defined. Fig. \ref{Supp:fig2}c shows the spectral density of the output intensity after propagation through a 76 mm long waveguide array (straight waveguides with equal separation of 24 $\mu$m). Up to 6 mW input power, which is the largest power used in the experiment, 76\% of the intensity (which is equivalent to the intensity found within the FWHM of a Gaussian) are found within a spectral range of 20 nm. The coupling variation due to SPM is therefore on the order of the intrinsic coupling uncertainty (see Fig. \ref{Supp:fig1}) and we can neglect this effect and use Eq. (1) in the main text to theoretically describe our experiments.

To relate the (experimental) time-averaged input power to the nonlinearity used in simulations, $g|\phi|^2$ (see Eq. (2) in the main text), we launch light into single waveguides in an array of straight waveguides with 24 $\mu$m separation. The low hopping constant ($J = 0.01$/mm) at this separation allows us to observe a change of the spatial output intensity pattern for input powers $\langle P_\text{in}\rangle$ below 1 mW, for which changes in the spectral shape are minimal (see also Fig. S2 c). At higher input powers above 1 mW, fitting results are distorted by the (temporal) tails of the pulse, which are governed by linear diffraction. For each input power, we find the corresponding $g|\phi|^2$ via fitting (least-square) the numerically-obtained output intensity pattern using Eq. (2) in the main text and including propagation losses. The dependence of $g|\phi(z=0)|^2$ on the input power is shown in Fig. \ref{Supp:fig2}d for five data sets, resulting in a value of $g$=(0.07 $\pm$ 0.01)/mm per mW input power.

\section{Obtaining solitons and linear stability analysis}

The instantaneous nonlinear eigenstates (e.g., solitons) in Fig. 3 in the main text and in the Supplementary Animation are obtained using a Newton iteration scheme. For that purpose, we use the FindRoot function in Mathematica \cite{mathematica} (with a precision of more than 15 digits). This method depends critically on the initial ansatzes. At high power ($g|\phi|^2/J^{\text{max}}=5$), we use six different ansatzes to evaluate the relevant soliton eigenstates belonging to six different branches. Three of the initial ansatzes are localized on a single site, while the other three ansatzes are localized between two sites with equal intensity in two neighboring sites. We then iteratively use the solitons at higher power as ansatzes for lower power. Additionally, we confirm the convergence by calculating the difference between the soliton obtained via the Newton Method, $\Phi_n^{\text{soliton}}$, and the eigenvector $\Phi_n^{\text{ev}}$ with the highest overlap, obtained from the nonlinear Hamiltonian $H_{\text{non-lin}} = H_{\text{lin}} -g |\Phi_n^{\text{soliton}}|^2$ in the following sense: $\sum_n |\Phi_n^{\text{soliton}}-\Phi_n^{\text{ev}}|^2 < 10^{-25}$.

Assuming a static (e.g., $z$-independent) Hamiltonian, a system prepared in an unstable nonlinear eigenstate does not guarantee a static evolution of the system as small fluctuations (even due to the limits of numerical precision) can amplify, and thus the soliton has finite lifetime. We test the stability of solitons using linear stability analysis, which indicates when solitons are linearly unstable. We follow \cite{RN2484} and start with a nonlinear eigenstate $\Phi^{(0)}(z)=e^{-i\Lambda z} \Phi^{(0)}$ for the nonlinear Schr\"odinger equation:
\begin{equation}\label{eq:S1}
	i \frac{d}{dz} \Phi = H \Phi - g |\Phi|^2 \Phi,
\end{equation}
where $H$ is the linear tight-binding Hamiltonian of the system, the parameter $g$ describes the strength of the nonlinearity, $\Phi$ is the (discrete) wavefunction and $\Lambda$ describes the eigenvalue of the eigenstate $\Phi^{(0)}$. Compared to Eq. (1) in the main text, we have dropped the subscripts, but it is clear that this equation should be understood as a tight-binding matrix equation. 

To test for stability, we take the following ansatz for a small perturbation around the solution: $\Phi= e^{-i\Lambda z}[\Phi^{(0)} + \epsilon (v(z)+iw(z))]$. Plugging this ansatz into Eq. (\ref{eq:S1}) and using the fact that $\Phi^{(0)}(z)$ solves the equation, we arrive (after some algebra) to first order in $\epsilon$ at the following two equations for the real and the imaginary parts of the perturbation:

\begin{equation}
\begin{aligned}
\frac{d}{dz} v =   (-\Lambda + H - 2g|\Phi^{(0)}|^2 + g(\Phi^{(0)})^2) w \equiv L_- w \\
\frac{d}{dz} w = - (-\Lambda + H - 2g|\Phi^{(0)}|^2 - g(\Phi^{(0)})^2 ) v \equiv -L_+ v \\
\end{aligned}
\end{equation}

These coupled equations are solved by separating the z-dependence via $v=e^{\kappa z} \overline{v}$ and $w=e^{\kappa z} \overline{w}$, which leads to: $\kappa \overline{w} = -L_+ \overline{v}$ and $\kappa \overline{v} =  L_- \overline{w}$. The values of $\kappa^2$ can now be calculated as eigenvalues of the matrix $-L_-L_+$. If $\kappa^2$ is positive, then $\kappa$ is real and the perturbations can build up exponentially. In contrast, if $\kappa^2$ is negative, then $\kappa$ is imaginary and the perturbations are oscillating waves that do not grow exponentially. In the latter case the soliton is linearly stable. In the main text, we identify solitons as stable if $\kappa^2 < 10^{-15}$.

\section{Confirming adiabatic behavior in the nonlinear system}

\begin{figure}[h!]
\includegraphics{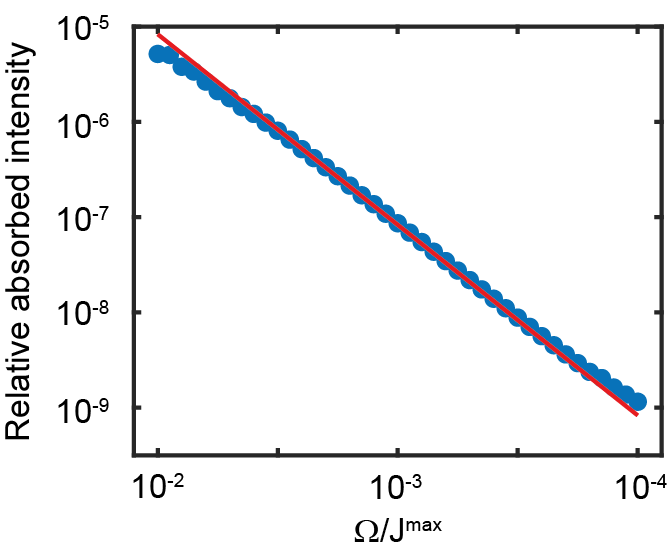}
\caption{\label{Supp:fig3} Adiabatic behavior in the nonlinear system. Absorbed intensity (relative to the total intensity) during one driving period in relation to the driving frequency. Blue circles are numerical values and the red line has a slope of -2. The parameters for the simulation are 180 sites with absorbing boundary conditions using 40 sites at each end, $d=24$ $\mu$m, $\bar{d}=2$ $\mu$m, $\alpha_0=-2\pi/12$ and $g|\phi|^2/J^{\text{max}}$=1.9.}
\end{figure}

In order to show that nonlinear Thouless pumps, similarly to linear pumps, have an adiabatic regime and therefore show perfectly quantized pumping in the adiabatic limit, we numerically evaluate the intensity radiated by the soliton during propagation. We use a system of 180 sites and numerically calculate the remaining intensity with absorbing boundary conditions on 40 waveguides at each end. Fig. \ref{Supp:fig3} illustrates that the absorbed intensity is approximately proportional to the square of the driving period $\Omega^{2}$ and therefore an adiabatic regime exists.



\newpage
\section{Quantized pumping of solitons bifurcating from all bands}
The pumped soliton solution presented in the main text bifurcates from the lowest band which has a Chern number of -1. Here, we show that there also exist soliton solutions bifurcating from higher bands (with Chern numbers of +2 and -1) which are pumped by the respective Chern number of the band from which they bifurcate. Fig. S4 displays the $z$ propagation over three full pumping cycles for solitons bifurcating from the lowest, the middle and the upper band at a power of $g|\phi|^2/J^{\text{max}}=0.2$, which are displaced by -1, +2 and -1 unit cells per pumping cycle, respectively. In contrast to traditional (linear) Thouless pumping, the projection of the soliton wavefunction onto the complete set of Bloch eigenstates (see insets in Fig. 4) clearly shows a non-uniform band occupation.

\begin{figure}[h!]
\includegraphics{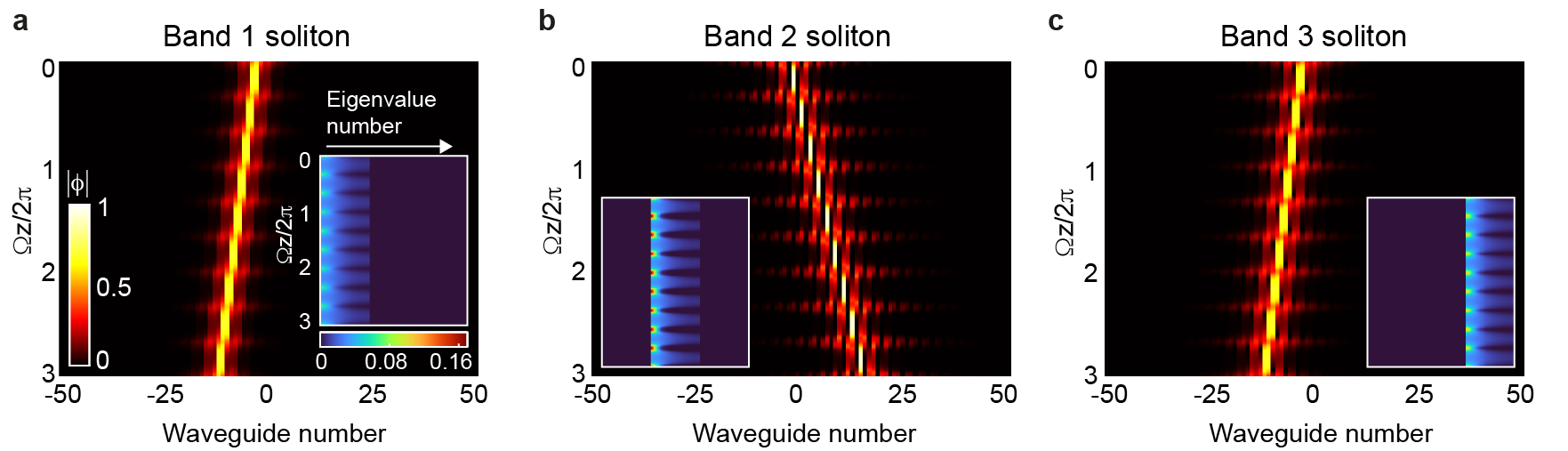}
\caption{\label{Supp:fignew} Quantized pumping of solitons bifurcating from all bands. \textbf{a,} Numerically calculated $z$-evolution over three full pumping cycles for an initial excitation of a soliton that bifurcates from the lowest band, showing a displacement of -1 unit cell per period. The inset shows the occupation of the soliton wavefunction $|\phi_{\text{soliton}}\rangle$, onto the Bloch eigenfunctions $|\phi_{\text{bloch}}\rangle$: $|\langle\phi_{\text{soliton}}|\phi_{\text{bloch}}\rangle|^2$. The projection coefficients are ordered with increasing energy eigenvalue of the Bloch functions. Note the almost exclusive but non-uniform occupation of the band from which the soliton bifurcates. \textbf{b,} Same as a, but for an initial excitation of a soliton that bifurcates from the middle band and which is displaced by +2 unit cells after each cycle. \textbf{c,} Same as a, but for an initial excitation of a soliton that bifurcates from the upper band and which is pumped by -1 unit cell after each cycle. Parameters for the system are chosen identical to Fig. 2, except for a lower power of $g|\phi|^2/J^{\text{max}}=0.2$ and $L=8 \cdot 10^5$ mm.}
\end{figure}

\section{Calculation of the Nonlinear Topological Invariant}
As mentioned in the main text, the topological invariant describing traditional (linear) Thouless pump is the Chern number, which is calculated using Bloch eigenstates. In the nonlinear case, translation invariance is broken and no notion of a band structure exists. Hence, a novel type of description is needed. Here, we calculate the Chern number of the pumped soliton a-posteriori, after the propagation is known. As the soliton is exponentially localized throughout the pump cycle, we can define an extended unit cell within which the soliton is entirely localized which then has translation symmetry. Fig. S4a shows such an extended unit cell, with 12 sites and four periods. The corresponding band structure, where the non-linearly induced potential is treated as a linear potential, is shown in Fig. 4b. Using Ref. \cite{RN2347} we numerically calculate the Chern number of the lowest band which describes the motion of the pumped soliton to be -1.

\begin{figure}[h!]
\includegraphics{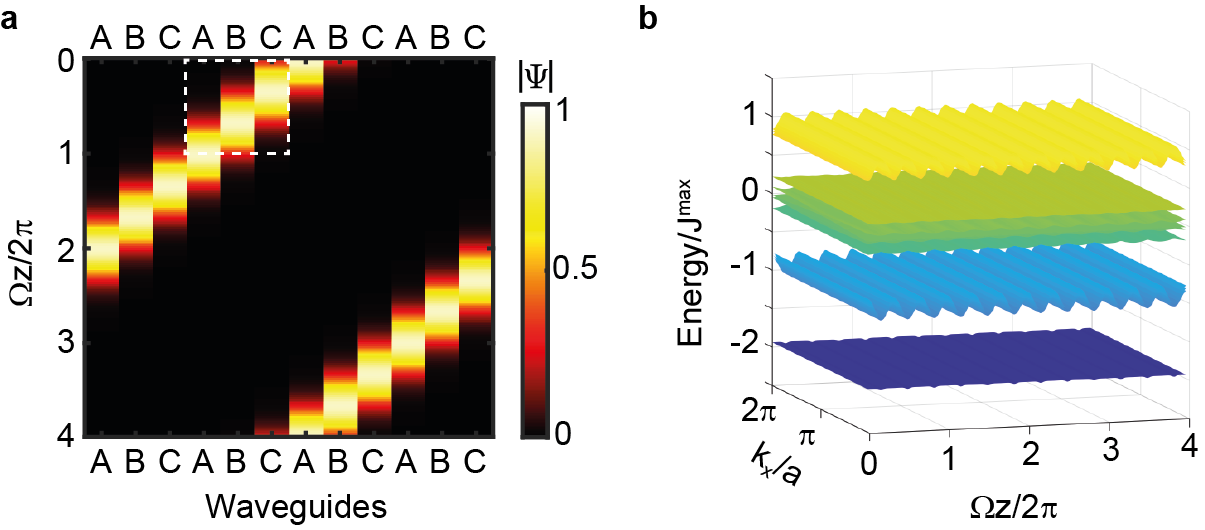}
\caption{\label{Supp:fig4} Calculation of the Chern number for the pumped soliton. \textbf{a,} Extended unit cell for a pumped soliton comprising of 12 sites and four periods. The white-dashed rectangle indicates the original unit cell (three sites, one period). \textbf{b,} Band structure with 12 bands corresponding to the extended unit cell from \textbf{a}, where the nonlinearly-induced potential is treated as a linear periodic potential. The clearly separated lowest band describes the motion of the pumped soliton and has a Chern number of -1. The parameter $a$ describes the spatial length of the unit cell.}
\end{figure}

\section{Description of the supplementary animation}
The supplementary animation shows the position of relevant soliton solutions over one period for increasing input power. It is directly related to Fig. 3 in the main text, which only shows the slices at four input power values. The animation shows additional unstable eigenstates which are not included in Fig. 2 to keep the figures in the main text uncluttered.

\bibliographystyle{ieeetr}

\bibliography{bibliography}